\documentstyle[epsfig,preprint,aps,prl]{revtex}
\tighten
\input epsf.sty

\begin{document}
\draft

\title{Bose-Einstein Condensation of Atomic Hydrogen}

\author{Dale~G.~Fried, Thomas~C.~Killian, Lorenz~Willmann,
David~Landhuis, Stephen~C.~Moss, Daniel~Kleppner, and Thomas~J.~Greytak}

\address{Department of Physics and Center for Materials Science and
Engineering,\\ Massachusetts Institute of Technology, Cambridge,
Massachusetts 02139}

\date{Accepted for publication in Physical Review Letters, October 2, 1998}

\maketitle

\begin{abstract}
We report observation of Bose-Einstein condensation of a trapped,
dilute gas of atomic hydrogen.  The condensate and normal gas are
studied by two-photon spectroscopy of the $1S$-$2S$ transition.
Interactions among the atoms produce a shift of the resonance
frequency proportional to density.  The condensate is clearly
distinguished by its large frequency shift.  The peak condensate
density is $4.8\pm1.1 \times 10^{15}~{\rm cm}^{-3}$, corresponding to
a condensate population of $10^9$ atoms.  The BEC transition occurs at
about $T=50~\mu$K and $n=1.8 \times 10^{14}~{\rm cm}^{-3}$.
\end{abstract}



The search for Bose-Einstein condensation in dilute atomic gases began
in earnest in 1978 \cite{greytakkleppnerproposal}, 
precipitated by the suggestion
of Stwalley and Nosanow \cite{sna76} that spin-polarized atomic
hydrogen should be an ideal candidate for the study of such extreme
quantum behavior.  Dilute gases of spin-polarized
H were first stabilized by Silvera
and Walraven \cite{swa80}.  Subsequent attempts to achieve BEC were
thwarted by recombination on the walls of the confinement cell
\cite{3bodyrecombrefs}.  In order to create the colder samples
necessary for further progress toward BEC, our group suggested
\cite{hes86} and demonstrated \cite{mds88} wall-free confinement and
evaporative cooling.  Evaporative cooling made it possible to cool
alkali metal atoms below the laser cooling limit and to achieve BEC in
dilute gases of Rb \cite{Aem95}, Na \cite{Dma95}, and Li \cite{Bst95}.
In this paper we report the achievement of BEC in H and a study of the
gas by high resolution spectroscopy.

Hydrogen condensates share many characteristics with condensates of
alkali metal atoms, but there are several notable differences.  Owing
to hydrogen's simplicity, properties such as interatomic potentials
and spin relaxation rates are well understood theoretically.  Compared
to those of other atoms, the $s$-wave scattering length, $a$, and
three-body loss rate are anomalously low.  As a result the condensate
density is high, even for small condensate fractions, and the elastic
collision rate is low, which retards evaporative cooling.  Because of
hydrogen's small mass, the BEC transition occurs at higher
temperatures than previously observed.  The cryogenic trap loading
technique used for H allows orders of magnitude more atoms to be
trapped, providing condensates that are much larger than yet achieved
in other systems.  Finally, high resolution two-photon spectroscopy
provides a new tool for studying condensates.

The starting point for the observations reported here is described in
the accompanying paper \cite{kfl98}: a gas of atomic hydrogen,
confined in a cylindrically symmetric magnetic trap, and cooled to
$120~\mu$K by evaporation over a magnetic field saddle\-point at one
end of the trap \cite{gre95}.  Below about $120~\mu$K this method of
forcing evaporation proved to be inefficient.  As explained by Surkov
{\em et al.} \cite{sws9496}, when the temperature is reduced and the
atoms settle into the harmonic region of the trap, the axial and
radial motional degrees of freedom become uncoupled.  For efficient
evaporation, the escape time for energetic atoms must be short
compared to the collision time.  However, the escape time for atoms
with high radial energy becomes very long when the motion uncouples,
and thus escape is blocked for this large class of energetic atoms.
By itself this would only slow evaporation, but in the presence of
losses due to dipolar relaxation \cite{hes86,mds88} phase space
compression ceases.

To overcome this problem, we turned to a different ejection technique,
based on spin resonance \cite{phm88}, first demonstrated for the
alkali metal atoms \cite{alkaliRF}.  An RF field causes transitions to
an untrapped magnetic sublevel wherever the trapping magnetic field
satisfies the resonance condition.  Atoms with high energy in any
degree of freedom are thus able to escape, assuring high evaporation
efficiency. As the RF frequency is decreased, the energy threshold for
ejection from the trap is lowered.  To implement RF evaporation in the
cryogenic environment, we developed a non-metallic trapping cell which
allows us to apply RF magnetic fields with strength up to $2\times
10^{-7}$~T and frequency up to $46~$MHz.  Typically we
switch to RF evaporation at a trap depth of 1.1~mK, corresponding to a
sample temperature of $120~\mu$K.

\begin{figure}
\vspace{0.5in}
\centering\epsfig{file=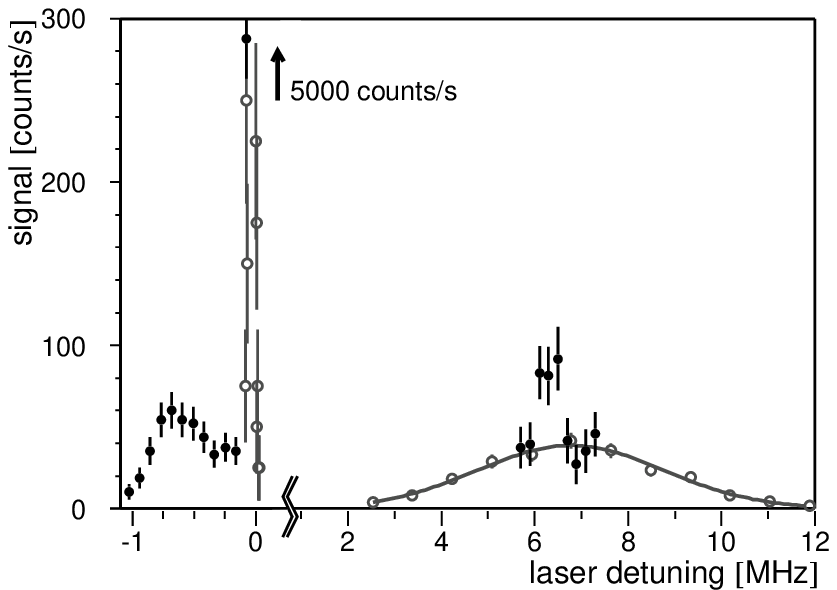,width=6in}
\vspace{.2in}
\caption{Composite $1S$-$2S$ two-photon spectrum of trapped
hydrogen. (Note change in frequency scale near the origin.)
$\circ$--spectrum of sample without a condensate; $\bullet$--spectrum
emphasizing features due to a condensate.  The intense, narrow peak
arises from absorption of counter-propagating photons by the normal
gas, and exhibits no first-order Doppler broadening.  The wide, low
feature on the right is from absorption of co-propagating photons.
The solid line is the recoil-shifted, Doppler-broadened, Gaussian
lineshape of the normal gas corresponding to $T=40~\mu$K.  The high
density in the condensate shifts a portion of the Doppler-free line to
the red.  The condensate's narrow momentum distribution gives rise to
a similar feature near the center of the Doppler-sensitive line. Zero
detuning is taken for unperturbed atoms excited Doppler-free.  All
frequencies are referenced to the $243$~nm excitation radiation.}
\label{overall.spec}
\end{figure}

We measure the temperature and density distribution of the gas by
two-photon spectroscopy of the $1S$-$2S$ transition \cite{cfk96}.
Two-photon excitation in a standing wave leads to two types of
absorption: Doppler-sensitive, due to absorption of co-propagating
photons, and Doppler-free, due to absorption of counter-propagating
photons.  In the first case the spectrum exhibits a recoil shift of
$\hbar k^2/2\pi m=6.7~$MHz and Doppler broadening that can be used to
measure the absolute temperature.  In the second case, Doppler-free
excitation results in a narrow, intense line.  These features are
shown in the composite spectrum, Fig.~\ref{overall.spec}.

When the sample is cooled below $50~\mu$K, additional features appear
in the spectrum, indicating the onset of BEC\@.  The Doppler-free
spectrum exhibits a new feature with a large red-shift.  As described
in the accompanying paper \cite{kfl98}, the $1S$-$2S$ transition
frequency is red-shifted by an amount proportional to the density.
Thus, this new feature indicates the formation of a region of high
density, as expected for a condensate.  Another new feature also
appears in the Doppler-sensitive spectrum, which probes the sample's
momentum distribution; the narrow momentum distribution of the
condensate appears as an additional feature in the middle of the wide,
recoil-shifted Doppler profile of the thermal cloud \cite{san93}.
This feature, whose intrinsic width should be given by the 
position-momentum uncertainty
relation, is broadened in these experiments by the high condensate
density.  As expected, it appears qualitatively similar to the
Doppler-free spectrum of the condensate.

\begin{figure}
\centering\epsfig{file=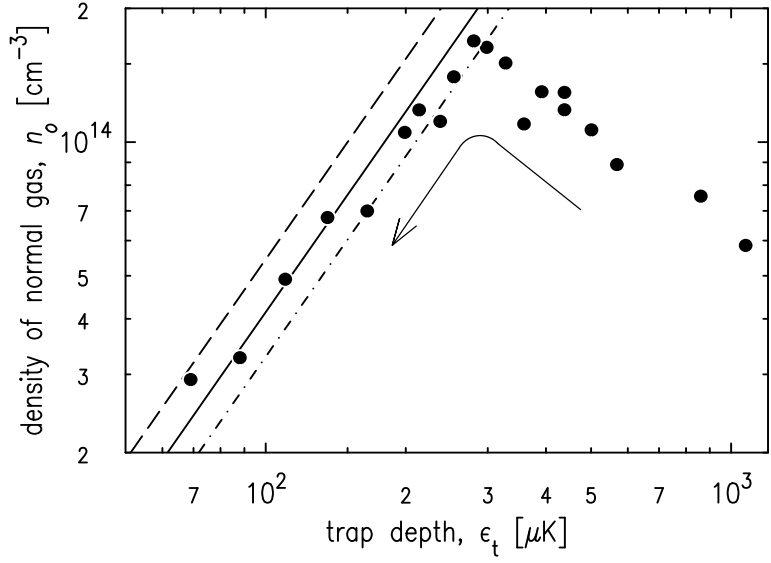,width=4in}
\vspace{.2in}
\caption{Density of non-condensed fraction of the gas as the trap
depth is reduced along the cooling path.  The density is measured by
the optical resonance shift {\protect \cite{kfl98}}, and the trap
depth is set by the RF frequency.  The lines (dash, solid, dot-dash)
indicate the BEC phase transition line, assuming a sample temperature
of (1/5th, 1/6th, 1/7th) the trap depth.  The scatter of the data
reflects the reproducibility of the laser probe technique and is
dominated by alignment of the laser beam to the sample.}
\label{phasespace}
\end{figure}

The onset of BEC is further confirmed by the trajectory of the normal
gas as it is cooled through phase space (see Fig.\ \ref{phasespace}).
The maximum density of the normal gas, $n_0$, is determined by
measuring the shift of the Doppler-free spectrum of the normal gas
\cite{kfl98}.  As the trap depth, $\epsilon_t$, is reduced from 1.1~mK
to $300~\mu$K, the density increases because the cooled atoms settle
into the lowest energy regions of the trap.  At a depth of $300~\mu$K
there is a change in the trajectory; it begins to {\em fall} with
temperature, indicating the onset of BEC.  For a sample in equilibrium
below the transition temperature, the maximum density of the normal
fraction is pinned at the critical density $n_c(T)=2.612 (2\pi m k_B
T)^{3/2}/h^3$; excess population is in the Bose condensate.  Although
it is not clear that our samples are in thermal equilibrium, this
portion of the trajectory approximates the BEC phase transition line.

The solid line in Fig.~\ref{phasespace} indicates the critical
density, $n_c(T)$, assuming that $\eta\equiv \epsilon_t/k_B T=6$.  The
parameter $\eta$ is set by a competition between cooling due to
evaporation and heating due to dipolar spin relaxation.  Dipolar
relaxation preferentially removes low energy atoms from regions of
highest density at the bottom of the trap.  The condensate has a high
density, and thus high loss rate. The value of $\eta$ is expected to
change from about 9 at higher temperatures in the absence of a
condensate to about 5 in the presence of a condensate at the lowest
trap depths.  The value $\eta=6$ was measured spectroscopically at a
trap depth $\epsilon_t=280~\mu$K \cite{doppler.data}.

\begin{figure}
\centering\epsfig{file=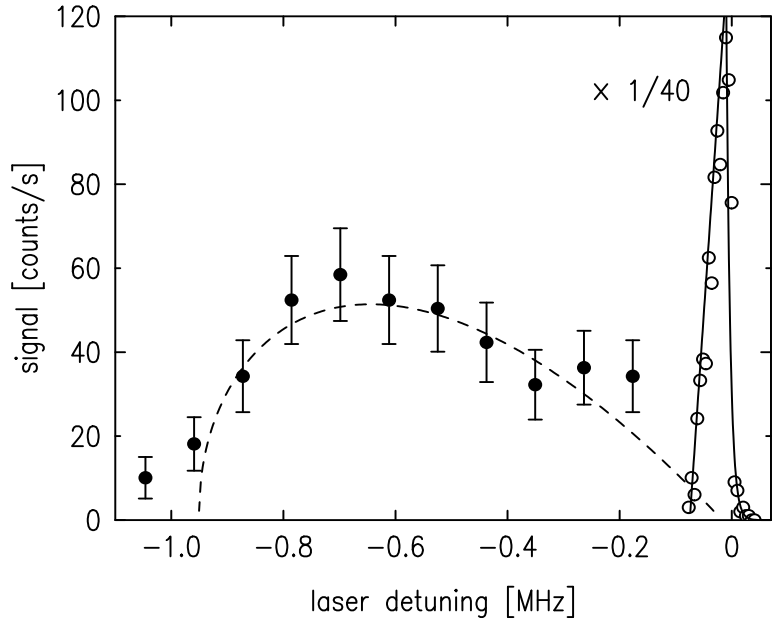,width=4in,angle=0}
\vspace{.2in}
\caption{Doppler-free spectrum of the condensate (broad feature) and
the normal gas (narrow feature).  The dashed line is proportional to
the number of condensate atoms at a density proportional to the
detuning, for an equilibrium density distribution with peak density
$n_p=4.8\times 10^{15}~{\rm cm}^{-3}$ in a parabolic trap.}
\label{condspec}
\end{figure}

Figure~\ref{condspec} is an enlarged view of the Doppler-free spectrum
of a Bose-condensed sample.  The high narrow feature is due to
excitation of the normal gas.  The condensate portion of the spectrum
extends to at least $-0.9$~MHz.  To extract the condensate density we
assume that the linear relation between frequency shift and density
for a normal gas found in Ref. \cite{kfl98} remains valid at high
densities.  For excitation out of a Bose-Einstein condensate, one
might expect that the disappearance of exchange effects would reduce
the frequency shift to one half of that for a normal gas at the same
density.  However, this reduction may not occur if the system
condenses into a group of low-lying states \cite{kss92}, in which case
the frequency shift is the same as for a normal gas \cite{lev98}.
Using this conservative assumption, for two-photon excitation,
$\Delta\nu_{1S-2S}=2 \Delta\nu_{243~{\rm nm}}=\chi\;n$ where
$\chi=-3.8\pm 0.8 \times 10^{-16}~{\rm MHz}~{\rm cm}^{3}$.  We derive
a peak condensate density of $n_p=4.8\pm1.1\times 10^{15}~{\rm
cm}^{-3}$. The spectrum depends on the distribution of densities in
the condensate, and thus the magnitude squared of the condensate
wavefunction.

The peak condensate density extracted from Fig.~\ref{condspec} is the
largest in this series of experiments, but several other spectra gave
comparable results.  From this peak density we compute the mean field
energy, the total number of condensate atoms, and the condensate
fraction.  We model our trap by the Ioffe-Pritchard potential
$V(\rho,z)=\sqrt{(\alpha \rho)^2 + (\beta z^2 + \theta)^2}-\theta$.
The parameters $\alpha$ and $\beta$ are calculated from the magnetic
coil geometry.  Because of the sensitivity of the axial bias,
$\theta$, to stray fields, we measure it using an RF ejection
technique; for these experiments $\theta/k_B=35\pm 2~\mu$K.  For small
displacements the radial oscillation frequency is
$\omega_\rho=\alpha/\sqrt{m \theta}=2\pi\times 3.90\pm0.11$~kHz, and
the axial frequency is $\omega_z=\sqrt{2\beta/m}=2\pi\times 10.2$~Hz.
The peak mean field energy $n_p \tilde{U}/k_B = n_p h^2 a/\pi m k_B =
1.9~\mu{\rm K}$ is much greater than the harmonic oscillator level
spacing, $\hbar \omega_r/k_B = 190~{\rm nK}$.

As the simplest model of the interacting gas, we assume the
Thomas-Fermi density profile for the condensate,
$n(\rho,z)=n_p-V(\rho,z)/\tilde{U}$.  The number of atoms in the
condensate is found by integrating the density over the volume of the
condensate.  The result is $N_c=16 \pi \sqrt{2}\: \tilde{U}^{3/2}\:
n_p^{5/2}/15\:\omega_\rho^2\: \omega_z \: m^{3/2} = 1.1\pm0.6\times
10^9$.  This condensate is $15~\mu$m in diameter and 5~mm in length.

Compared to alkali metal atoms, the maximum achievable equilibrium
condensate fraction for hydrogen is expected to be small \cite{hks93}.
Hydrogen's small $s$-wave scattering length dramatically reduces the
elastic collision rate in the thermal cloud, and thus the evaporative
cooling rate.  The high density in a condensate can drastically
increase the heating due to dipolar relaxation; the resulting
heating-cooling balance limits the condensate fraction.

The condensate fraction, $f\equiv N_c/(N_n+N_c)$, is obtained by
comparing the integrated area of the spectra of the condensate and
normal gas after accounting for the partial ($\sim 10$\%) illumination
of the normal gas by the laser beam.  A condensate fraction
$f=5_{-2}^{+4}\%$ is obtained, with the dominant uncertainty coming
from the relative size of the laser beam and thermal cloud.
Alternatively, the condensate fraction may be estimated by comparing
the condensate population, $N_c$, to the calculated normal gas
population, $N_{n}$, which is simply the Bose occupation function
weighted by the density of states, integrated over the trapped states.
The trap parameters $\alpha$ and $\beta$ cancel in the ratio
$N_c/N_{n}$, and the remaining parameters, $T$, $n_p$, and $\theta$,
are measured.  We use the Doppler broadened spectrum to measure the
temperature at $45\pm5~\mu$K.  Since this thermal energy scale is much
larger than the peak mean field energy, perturbations to the density
distribution of the normal gas due to the condensate may be neglected.
The resulting fraction is $f=6_{-3}^{+6}\%$.  Both methods for
determining $f$ assume thermal equilibrium, which may not be the case
because the mean collision time is comparable to the axial oscillation
period.

Note that if we had assumed that the frequency shift in the condensate
is only half as large as for a normal gas at the same density, as one
would expect for a condensate in a single quantum state, the density
extracted from the spectrum would be twice as large. This would imply
that $N_c=6\times 10^9$ and yield an unreasonably high condensate
fraction of 25\%.

We have followed the time evolution of individual condensates
by recording series of spectra in rapid succession. 
As expected, when the condensate decays the spectrum narrows,
indicating a reduction in density, and becomes less intense,
indicating a reduction in condensate population.  The lifetime is
observed to be about 5~s. The condensate decays rapidly by dipolar
relaxation, but is simultaneously replenished by the normal gas.  The
laser probe artificially shortens the lifetime by evaporating He from
optical surfaces; the streaming He knocks atoms out of the trap.

\begin{figure}
\centering\epsfig{file=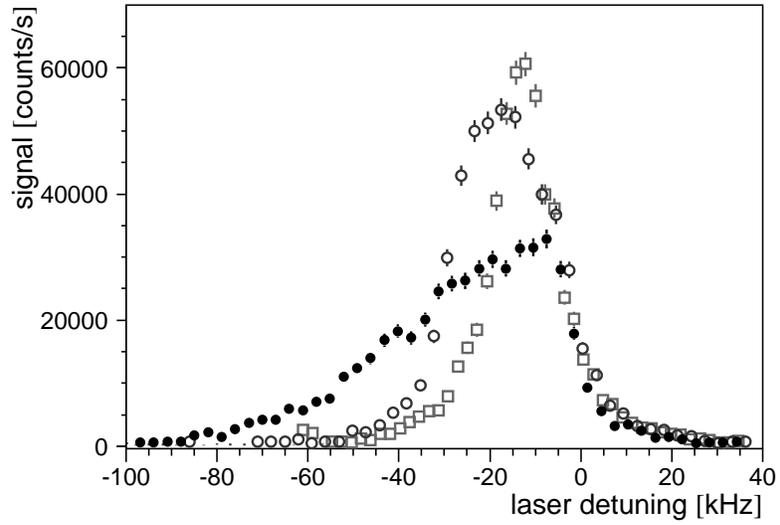, width=4in}
\vspace{.2in}
\caption{Doppler-free spectrum of normal fraction above and below the
onset of BEC\@.  The symmetric spectrum (above $T_c$, open symbols)
suddenly becomes asymmetric (filled symbols) when the condensate
forms.  Temperatures for the three spectra are about $120~\mu$K (open
squares), $53~\mu$K (open circles), $44~\mu$K (filled circles).}
\label{asymspec}
\end{figure}

We find that the presence of the condensate significantly distorts the
Doppler-free spectrum of the normal gas, even though the condensate
volume is only about $10^{-3}$ of the volume of the thermal cloud.
Figure~\ref{asymspec} shows spectra taken above and below the onset of
BEC\@.  The spectra for samples without condensates show the shift and
lineshape due to the inhomogeneous density.  The lines are roughly
symmetric, and are consistent with calculations assuming thermal
equilibrium.  Below the transition, however, a pronounced
asymmetry appears, showing spectral weight at frequency shifts much
larger than expected for the maximum density in the normal gas,
$n_c(T)$.  The origin of this effect is not yet understood.

The signal-to-noise ratio in these experiments is limited by the light
collection solid angle, about $2\times 10^{-2}$~ster.  Improvements
should allow the spectroscopic observation technique to provide
information which is complementary to that obtained by the spatial
imaging techniques used in other experiments; the Doppler-sensitive
spectrum constitutes an {\em in situ} direct measure of the momentum
distribution.

Note: Observation of quantum-degeneracy in a 2-dimensional hydrogen
system has been reported recently\cite{svy98}.

Many people played important roles in the MIT work on spin-polarized
hydrogen over the years.  We especially wish to acknowledge those who
contributed directly to the trapping and spectroscopy experiments:
Claudio L. Cesar, John M. Doyle, Harald F. Hess, Greg P. Kochanski,
Naoto Masuhara, Adam D. Polcyn, Jon C. Sandberg, and Albert I. Yu.
We thank Wolgang Ketterle and Leonid Levitov for helpful discussions.

This research is supported by the National Science Foundation and the
Office of Naval Research.  The Air Force Office of Scientific Research
contributed in the early phases.  L.W. acknowledges support by
Deutsche Forschungsgemeinschaft.  D.L. and S.C.M. are grateful for
support from the National Defense Science and Engineering Graduate
Fellowship Program.


\end{document}